\documentclass[a4paper]{jpconf}
\usepackage{graphicx,color}
\begin{document}

\title{Atomic physics experiments with trapped and cooled highly charged ions}

\author{H.-J. Kluge$^{1,2}$, W. Quint$^{1,2}$ and D.F.A. Winters$^{1}$}

\address{$^{1}$Gesellschaft f\"ur Schwerionenforschung (GSI) mbH, Atomic Physics Division,
Planckstrasse 1, D-64291 Darmstadt, Germany}

\address{$^{2}$Universit\"at Heidelberg, Physikalisches Institut, Philosophenweg 12,
D-69120 Heidelberg, Germany}

\ead{J.Kluge@gsi.de}

\begin{abstract}
Trapping and cooling techniques have become very important for
many fundamental experiments in atomic physics. When applied to
highly charged ions confined in Penning traps, these procedures
are very effective for testing quantum electrodynamics in extreme
electromagnetic fields produced by heavy highly charged ions such
as uranium U$^{91+}$. In addition, fundamental constants or
nuclear ground state properties can be determined with high
accuracy in these simple systems. Finally, by studying a single
trapped radioactive ion, its nuclear decay can be studied in
detail by observing the disappearance of the signal of the mother
and the appearance of that of the daughter isotope. Such
experiments on highly charged ions at extremely low energy will
become possible by the HITRAP facility which is currently being
built up at GSI. Also the future Facility for Antiproton and Ion
Research (FAIR) will be briefly described which is expected to be
operational by 2014.
\end{abstract}

\section{Why use highly charged ions?}
Quantum electrodynamics (QED) is the most precisely investigated
theory in physics \cite{BEI00}. It describes the interaction of
electric charges by exchange of photons, and serves as a basis for
many other existing field theories. Experimental studies have been
carried out with extremely high precision of up to $10^{-14}$,
always consistent with the QED predictions. For a few simple
systems, the experimental accuracy is matched by nearly equally
exact theoretical calculations, which shows the power of the
underlying mathematical framework. Discrepancies between theory
and experiment are nowadays often caused by insufficient knowledge
of the nuclear size and its structure. Even fundamental constants
are known to limit the predictive power of QED to a certain
extent. This, however, allowed high-accuracy experiments to
provide the most precise values for the fine-structure constant
$\alpha$ \cite{GAB06a} and the mass of the electron $m_e$
\cite{FLO04}.

For such precision studies, simple systems have to be
investigated, consisting of just an atomic nucleus and a few
electrons at most. Few-electron systems with a heavy nucleus are
particularly important because they allow one to access the regime
of extreme electromagnetic fields ($\approx 10^{16}$ V/cm) within
a range of distances of the order of the electron's Compton
wavelength. This is also the region where perturbative QED,
although being very precise, breaks down \cite{MOH98}. The energy
contained in these fields is also very close to that required for
the spontaneous creation of an electron-positron pair out of the
vacuum (Schwinger-limit).

Highly charged ions (HCI), combining very strong static fields and
a simple electronic structure, are ideal testing grounds for these
high-field QED investigations. HCI can readily be produced and
stored in traps and rings. Sensitive techniques, such as mass or
laser spectroscopy in storage rings, ion traps, and electron beam
ion traps, have been developed to further advance this field.
However, experiments with heavy HCI have not (yet) matched the
accuracy of experiments performed with hydrogen or simple light
systems, {\it e.g.}, the free electron or positronium.

HCI are either produced at once by impact of an energetic ion beam
on a `stripper target', or stepwise by electron impact ionisation
of trapped ions. Presently, the only way to produce sufficient
amounts of HCI (up to U$^{91+}$) is by using relativistic ion
beams (about 1 GeV per nucleon) in combination with a stripper
target, as being done at GSI.

\begin{figure}[b]
\begin{center}
\includegraphics[width=10cm]{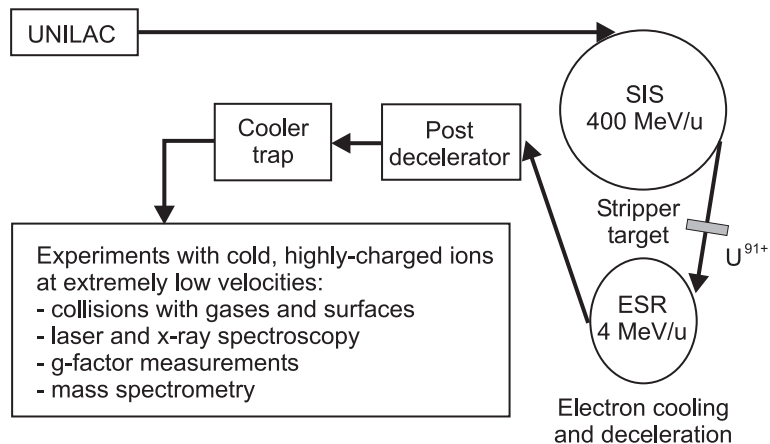}
\end{center}
\caption{\label{fig1}Schematic of the HITRAP project:
deceleration, trapping, and cooling of highly charged ions for
many key atomic physics experiments.}
\end{figure}

\section{Experiments with trapped and cooled highly charged ions}
In a Penning trap, charged particles are confined in the axial or
$z$-direction by an electric quadrupole field created by applying
static potentials ($\sim 10$ V) to the trap electrodes, which
typically consist of two endcaps and one ring electrode
\cite{BLA06}. In the radial or $xy$-plane, confinement is
generally achieved by use of a superconducting magnet which
provides a strong, stable, and homogeneous static magnetic field.
The motions that the trapped particles undergo are all simple
harmonic oscillations and are very well under control. The motion
in the $z$-direction is an oscillation with the axial frequency
$\omega_z$, and is the simplest motion to cool and to detect. In
the radial plane there are two circular motions: a fast motion
with the modified cyclotron frequency $\omega_+$, and a slow drift
motion about the centre of the trap with the magnetron frequency
$\omega_-$. The radiofrequency or Paul trap is another type of
trap for charged particle confinement, which is very often used,
but its description is beyond the scope of this paper.

Trapped (charged) particles form a point-like source and generally
have small amplitudes of oscillation. Once trapped, the particles
can easily be manipulated, ($q/m$)-selected, accumulated, and even
polarized ({\it e.g.}, by optical pumping). A number of cooling
techniques exist, which, besides removing the particles' kinetic
energy and thus reducing Doppler effects, also improve particle
manipulation, selection, and the resolution of the measurement.
Such cooling techniques are laser cooling, sympathetic cooling,
evaporative cooling, resistive cooling, and electron cooling. Cold
ensembles of trapped particles can also be formed into dense
crystalline ion structures which could, for example, lead to an
enhanced luminosity for reaction studies. Since confinement times
of trapped particles can indeed be very long (up to months), there
is an extended observation and manipulation time. Thus metastable
states can simply be eliminated by waiting, and a backing-free
sample for decay studies can be obtained. This also makes traps
very suitable for effective studies of rare species. Further
applications are bunching, charge breeding, and post-acceleration.
All the above mentioned methods lead to drastically increased
efficiency, accuracy and sensitivity. In principle, the highest
accuracy can be obtained when only a single (cold, highly charged)
ion is used. Trapped highly charged ions have been successfully
used for high-accuracy mass measurements in Seattle and Stockholm,
for producing and studying ionic crystals in Livermore, and for
the determination of the $g$-factor of the bound electron in
hydrogen-like ions in a GSI-Mainz cooperation (see below).

In order to further increase the accuracy, the Highly charged Ion
TRAP (HITRAP) facility is presently being built at GSI, and will
be operational in 2008 \cite{KLU05}. Briefly, stable or
radioactive HCI at relativistic velocities (400 MeV/u) are
decelerated (and cooled) in the experimental storage ring (ESR)
down to 4 MeV per nucleon, and injected into the HITRAP facility.
Here, they will be further decelerated by a linear decelerator and
a radiofrequency quadrupole structure (RFQ), and injected into a
cooler Penning trap, where a temperature of 4 K is reached by
electron and resistive cooling. The cold HCI are then transferred
at low energies (5 keV per charge) through a beamline to different
experimental setups. The planned unique experiments are being
prepared by several international groups (GANIL, Groningen, GSI,
Heidelberg, Krakow, London, Mainz, Stockholm, and Vienna) within
the EU-funded HITRAP project. A schematic of HITRAP is shown in
figure~\ref{fig1}, together with a list of some planned atomic
physics experiments. Below, two such experiments are described in
some more detail.

\begin{figure}[b]
\begin{center}
\includegraphics[width=10cm]{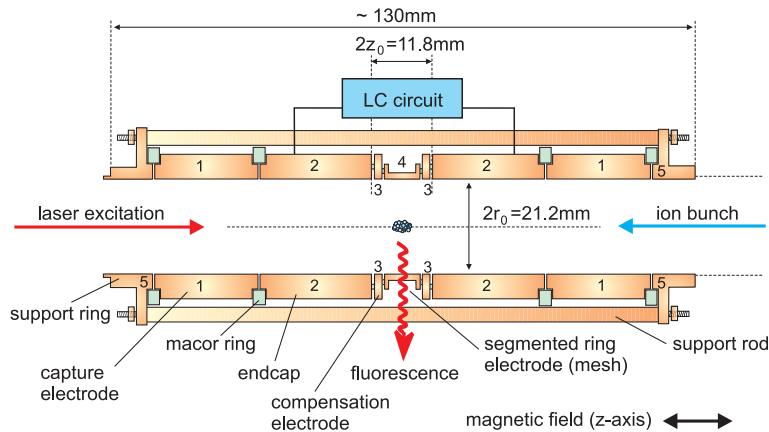}
\end{center}
\caption{\label{fig2}Schematic of the Penning trap that will be
used to measure ground-state hyperfine splittings in highly
charged ions. The ions come from the right, are trapped, cooled
and compressed, and are excited by a laser beam from the left. The
emitted fluorescence is detected through the optically transparent
ring electrode.}
\end{figure}

\subsection{Laser spectroscopy}
The ground-state hyperfine splitting (HFS) of heavy H-like ions
increases with the nuclear as $Z^3$ and enters the
laser-accessible regime for heavy elements such as
$^{207}$Pb$^{81+}$ or $^{209}$Bi$^{82+}$. The $1s$ HFS in these
H-like systems has been studied at the ESR storage ring at GSI by
collinear laser spectroscopy \cite{KLA94,SEE98}, and at the
SuperEBIT in Livermore \cite{CRE96,CRE98,BEI01,BEI03a}. There also
exist two measurements of the $2s$ ground state HFS in Li-like
bismuth ($^{209}$Bi$^{80+}$). A direct measurement \cite{BOR00}
was carried out at the ESR, but unfortunately no resonance could
be observed at the predicted value of $\approx 1554$ nm
\cite{SHA01}. The reason is unknown. An indirect measurement
\cite{BEI98} was performed in the SuperEBIT and yielded a value of
$\approx 1512$ nm, but the error in the measurement was rather
large ($\approx 50$ nm). From a comparison of the HFS of a H- and
a Li-like system, the nuclear effects ({\it e.g.}, Bohr-Weisskopf)
cancel out to a large extent, and the QED effects can be
determined within a few percent accuracy \cite{SHA01,WIN06}.
Furthermore, if the nuclear magnetic moments $\mu$ are not very
well known, HFS measurements in few-electron systems will also
allow for a determination of $\mu$ and can thus help to improve
the knowledge of nuclear properties.

For these laser experiments at HITRAP a cylindrical open-endcap
Penning trap \cite{VOG05} will be mounted inside the cold (4 K)
bore of a superconducting magnet (RETRAP \cite{GRU05}) with radial
and axial access. RETRAP was previously used to trap Xe$^{44+}$
ions, produced by an electron beam ion trap, and to study the
formation of ion crystals by sympathetic cooling with laser-cooled
Be$^+$ ions \cite{GRU05}. The planned HFS laser measurements at
HITRAP will take place within an international collaboration
formed by GSI, Imperial College London, TU Darmstadt, Lawrence
Berkeley and Livermore Labs, and Texas A\&M University.

\begin{figure}[b]
\begin{center}
\includegraphics[width=6cm]{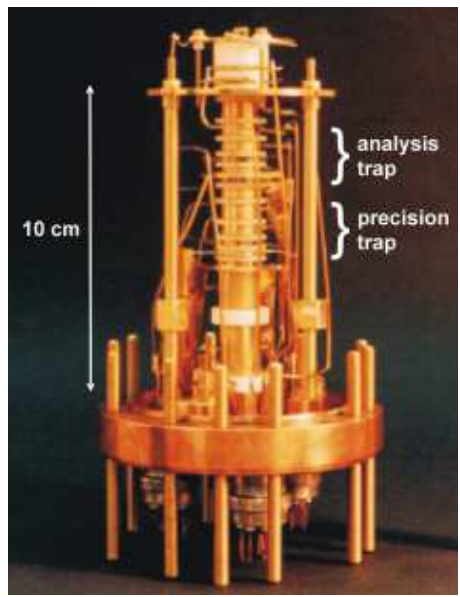}
\end{center}
\caption{\label{fig3}Photograph of the GSI-Mainz $g$-factor Penning trap.}
\end{figure}

\subsection{Mass spectrometry}
The masses of stable or radioactive nuclides can be measured by
Penning traps with very high accuracy and single-ion sensitivity
\cite{LUN03,BLA06}. Since the cyclotron frequency of an ion in a
Penning trap increases with the charge state, HCI provide better
resolution and potentially also higher accuracy than singly
charged ions. The SMILETRAP group at Stockholm has pioneered the
use of HCI in ion traps for mass spectrometry
\cite{JER91,FRI06,NAG06} and measured with an uncertainty close to
$10^{-10}$ the masses of several ions, which are important for
fundamental tests of QED or double-beta decay. Using the mass of
$^{12}$C$^{6+}$, the masses of singly charged stable ions have
been measured with an accuracy of about $10^{-10}$ in Seattle by
Van Dyck {\it et al.} \cite{DYC06}. The group of Gabrielse
achieved a similar accuracy in the case of the proton-antiproton
comparison at CERN \cite{GAB06}. An even better accuracy of about
$10^{-11}$ was obtained at MIT by Pritchard {\it et al.}
\cite{THO04} with a mass spectrometer now at Florida State
University \cite{RED06}. The masses of singly charged
radionuclides have been measured with an accuracy in the range of
$10^{-7}$ to $10^{-8}$ and better by Penning trap mass
spectrometers installed at ISOLDE/CERN \cite{HER01}, Argonne
\cite{SAV06}, Jyv\"askyl\"a \cite{ERO06}, MSU \cite{BOL06a}, and
GSI \cite{RAU06}. HCI for mass spectrometry will be used at TITAN
at ISAC/TRIUMF, Vancouver \cite{DIL06}, LEBIT/MSU \cite{BOL06b},
HITRAP/GSI \cite{HER06a}, ISOLTRAP/ISOLDE \cite{HER06b},
MAFFTRAP/Munich \cite{HAB06}, and at MATS/Darmstadt \cite{BLA06}.

A sensitive test of bound-state QED in strong fields is the
measurement of the $g$-factor of the electron bound to a nucleus
in a hydrogen-like ion. The ratio of the bound-electron
($g_{bound}$) to the free-electron g-factor ($g_{free}$) can be
expressed to leading order in $Z \alpha$ as
\begin{equation}
\frac{g_{bound}}{g_{free}} \approx 1- \frac{1}{3} (Z \alpha)^2 + \frac{1}{4 \pi} \alpha (Z \alpha)^2,
\end{equation}
where the second term stems from Dirac theory and the third from bound-state QED.

Within the Mainz-GSI collaboration, the $g$-factor of the bound
electron in $^{12}$C$^{5+}$ \cite{HAF00} and $^{16}$O$^{7+}$
\cite{VER04} has been obtained via spin-flip measurements of a
single cold (4 K) ion. The setup consists of two Penning traps
\cite{HAF03}, {\it i.e.} a `precision trap' and an `analysis
trap', in one superconducting magnet. Figure~\ref{fig3} shows a
photograph of the Penning trap. The $g$-factor can be obtained
from measurements of the cyclotron frequency $\omega_c=qB/M_{ion}$
and the Larmor frequency $\omega_L=geB/(2m_e)$, as it can be
expressed as
\begin{equation}
g=2 \left( \frac{q}{e} \right) \left( \frac{m_e}{M_{ion}} \right) \left(\frac{\omega_L}{\omega_c} \right),
\end{equation}
where $m_e,e$ and $M_{ion},q,$ are the mass and charge of the
electron and the ion, respectively. The three oscillation
frequencies of an ion inside a Penning trap can be measured
independently and with high accuracy, using resonant circuits with
high quality factors. The polarisation of the electron spin was
100\%, simply because only one ion was used. Inside the precision
trap $\omega_c$ is determined, and a spin-flip may be induced by
microwave irradiation at a frequency close to the spin precession
frequency $\omega_L$. Detection of a spin-flip takes place after
transporting the ion to the analysis trap. Here, a significant
inhomogeneity of the magnetic field, produced by a nickel ring,
makes the $z$-motion of the ion sensitive to the spin direction.
The frequency of the $z$-motion is detected via the image charges
induced in the endcap electrodes. After analyzing the spin
direction, the ion is transported back and the next measurement
cycle begins.

\begin{figure}[t]
\centering
\begin{minipage}{0.4\textwidth}
  \includegraphics[width=1\textwidth]{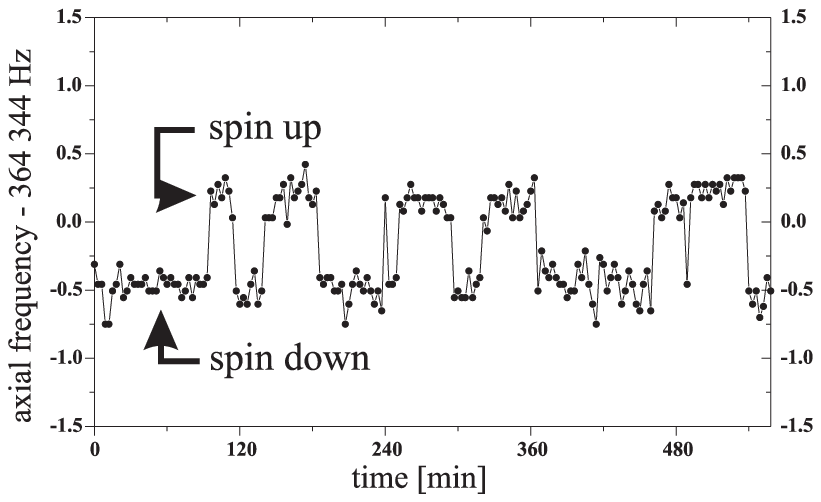}
\end{minipage}
\hspace{1.0cm}
\begin{minipage}{0.4\textwidth}
  \includegraphics[width=1\textwidth]{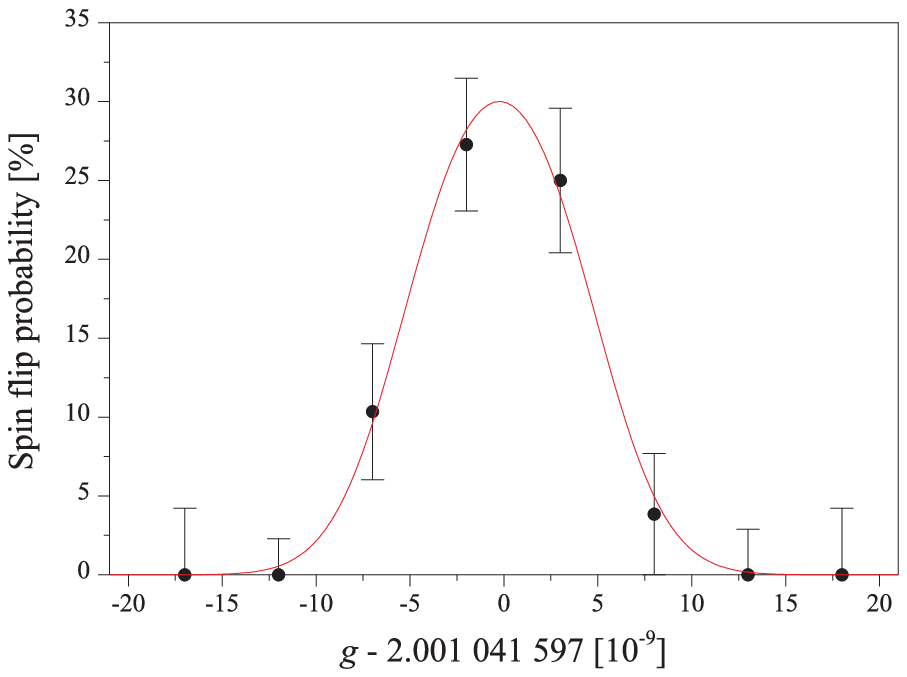}
\end{minipage}
\caption{\label{fig4}Result of the $g$-factor measurement of the
bound electron in $^{12}$C$^{5+}$ \cite{HAF00}. The left panel
shows individual spin-flips, the right panel the Larmor
resonance.}
\end{figure}

Figure~\ref{fig4} shows the experimental result obtained for a
$g$-factor measurement of a single $^{12}$C$^{5+}$ ion. The
abscissa indicates the measured Larmor precession frequency of the
bound electron which directly yields, after dividing by the
simultaneously measured ion cyclotron frequency, the $g$-factor of
the bound electron. The experimental $g$-factors of the
hydrogen-like ions $^{12}$C$^{5+}$ \cite{HAF00} and
$^{16}$O$^{7+}$ \cite{VER04} agree within the uncertainties, which
are dominated by the accuracy of the electron mass, with the
theoretical ones \cite{BEI00}. The $g$-factor found for the $1s$
electron in these systems enabled a test of bound-state QED on a
0.25\% level. Because of the good agreement of these $g$-factors
with theory, the data could be used to determine the electron mass
four times more accurately than the previously accepted value
\cite{BEI02}. The achieved experimental accuracy $\delta m_e/m_e$
of this measurement was as good as $6 \times 10^{-10}$.

In table 1, a comparison is made between a mass measurement using
a singly charged ion in a Penning trap and one employing a highly
charged ion. From the table it is clear that, for the same
isotope, a HCI with a charge $q$ of, for example, 50 already leads
to an improved mass resolution of nearly two orders of magnitude.
Higher charge states ({\it i.e.} $q=92$ for the case of uranium)
and longer observation times (for example, $T_{obs}>=10$ s) will
improve the mass resolution even further and will make it possible
to reach a mass measurement accuracy of $10^{-11}$ or better with
highly charged ions. This would allow one to measure the $1s$ Lamb
shift in hydrogen-like uranium by weighing with an accuracy better
than by x-ray spectroscopy.

\begin{table}[t]
\caption{Comparison of the accuracy of mass measurements as
obtained for a singly and for a highly charged ion with mass
number $A=100$ in a magnetic field of $B=6$ T. $\nu_c$ is the
cyclotron frequency, $T_{obs}$ the observation time of a
measurement, $R$ is the resolving power given by the ratio $\nu_c
/ \delta \nu_c$, and $\delta m/m$ is the mass uncertainty.}
\centering
\begin{tabular}{|ll|ll|}
\hline
singly charged ion &  & highly charged ion & \\
\hline
$q=1$ & $\nu_c=1$ MHz & $q=50$ & $\nu_c=50$ MHz \\
$T_{obs}=1$ s & $\delta \nu_c=1$ Hz & $T_{obs}=1$ s & $\delta \nu_c=1$ Hz \\
$R=10^6$ & $\delta m/m \approx 10^{-8}$ & $R=5 \times 10^7$ & $\delta m/m \approx 2 \times 10^{-10}$ \\
\hline
\end{tabular}
\end{table}

\section{FAIR - the future Facility for Antiproton and Ion Research}
In the more distant future, HITRAP will be a component of the
Facility for Low-Energy Antiproton and Ion Research (FLAIR) at the
future international accelerator facility FAIR \cite{FAI06}.
There, HITRAP will not only provide low-energy highly charged
stable or radioactive ions for experiments of the SPARC
Collaboration, but also low-energy antiprotons. Furthermore, the
FAIR facility will provide highest intensities of both stable and
radioactive ion beams and energies up to 34 GeV per nucleon. At
such energies, the HCI generate electric and magnetic fields of
exceptional strength and ultra-short duration.

\begin{figure}[t]
\begin{center}
\includegraphics[width=12cm]{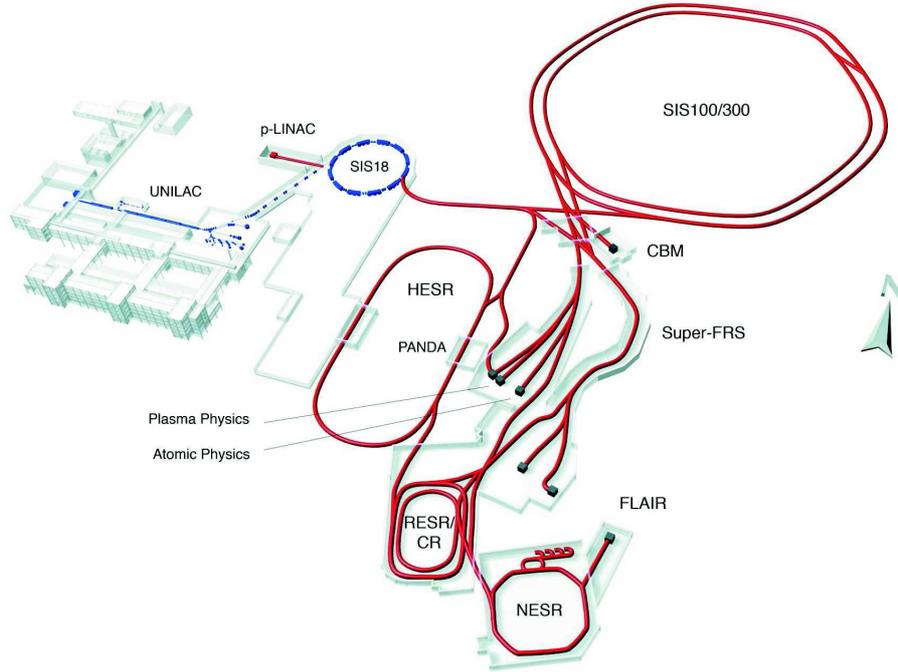}
\end{center}
\caption{\label{fig5}Schematic of the future facility for
antiproton and ion research (FAIR) at GSI.}
\end{figure}

\subsection{Stored Particle Atomic Research Collaboration (SPARC)}
The new FAIR facility has key features that offer a range of new
opportunities in atomic physics research and related fields, which
will be exploited by the atomic physics Collaboration SPARC
\cite{STO05}. In particular, at FAIR the Super Fragment Separator
(SFRS) will provide a rich spectrum of radionuclides that are not
available at any other facility. The high intensity of secondary
beams produced at the SFRS will make it possible to extract
decelerated radioactive ion beams from the New Experimental
Storage Ring (NESR) and to decelerate them for trap experiments
with sufficient intensity at HITRAP. Therefore, the physics
programme of HITRAP at FLAIR can be extended to novel experiments
with trapped radioactive ions and, of course, with trapped
antiprotons. Trapped radioactive ions in high charge states may
reveal a completely new domain for fundamental interaction studies
and for experiments at the borderline between atomic and nuclear
physics.

Moreover, the manipulation of trapped radioactive ions with laser
light opens up possibilities to study questions of the Standard
Model of fundamental interactions in a unique way at HITRAP. By
optical pumping within the hyperfine levels of the ground state,
the nuclear spins of radioactive nuclides can be polarized with
high efficiency. The detection of the asymmetry of beta decay, for
example, will allow one to explore the vector/axial-vector
(VA)-structure of the weak interaction and to set limits for the
masses of heavy bosons, which are not included in the Standard
Model.

Direct mass measurements on unstable nuclides with ultra-high
accuracy up to $\delta m/m \approx 10^{-11}$ are another class of
investigations which become possible at the HITRAP facility at
FLAIR. Such an accuracy would allow one to determine the 1s-Lamb
shift of $U^{91+}$ with an accuracy of $\delta mc^2 \approx 2 eV$,
better than presently possible by X-ray spectroscopy \cite{GUM05}.
If the QED calculations are found to be correct, nuclear charge
radii also of unstable nuclides can be determined. For a general
exploration of the nuclear mass surface in the chart of nuclei an
accuracy in the mass determination of $\delta m/m \approx 10^{-6}$
to $10^{-7}$ is in general sufficient as planned by isochronous or
Schottky mass spectrometry experiments at the NESR storage ring.
However, in some cases, like double-beta decay or tests of the
unitarity of the Cabibbo-Kobayashi-Maskawa matrix, a much higher
accuracy is required, which is possible by use of HCI stored in a
Penning trap at HITRAP \cite{HER06a} and at MATS \cite{BLA06}.

\subsection{Facility for Low-Energy Antiproton and Ion Research (FLAIR)}
The planned FLAIR facility will be the most intense source of
low-energy antiprotons world-wide \cite{WID05}. The beam intensity
of extracted low-energy antiprotons will be two orders of
magnitude higher in the FLAIR facility compared to the Antiproton
Decelerator (AD) at CERN. Hence we anticipate that experiments
with trapped antiprotons will be performed at the FLAIR facility,
which are currently impossible anywhere else due to intensity
reasons. A possible highlight in the field of low-energy
antimatter research would be the first direct experimental
investigation of the gravitational interaction of antimatter which
has never been attempted up to now. Such investigations could be
performed on ultra-cold antihydrogen atoms \cite{WAL04} which are
produced by recombining trapped antiprotons with positrons in a
so-called nested Penning trap. The effect of gravity on antimatter
is an important issue for the development of quantum theories of
gravity.

\ack We would like to thank all our colleagues and collaborators
for letting us use their work for this report. We acknowledge
support by the European Union, the German Ministry for Education
and Research (BMBF) and the Helmholtz Association (HGF).

\end{document}